\documentclass[12pt,a4paper]{article} 
\usepackage{relsize}
\usepackage[top=1.70in, bottom=1.70in, left=1.25in, right=1.25in]{geometry}
\usepackage{graphicx}
\usepackage{amsmath}
\usepackage{mathtools}
\usepackage{relsize}
\usepackage{graphicx} 
\usepackage[caption=false]{subfig} 
\usepackage{float}
\providecommand{\e}[1]{\ensuremath{\times 10^{#1}}}
\usepackage{multirow}
\usepackage[toc,page]{appendix}
\usepackage[utf8]{inputenc}
\begin{document}

\begin{titlepage}
	\centering
	{\scshape\LARGE PSI \par}
	\vspace{1cm}
	{\huge\bfseries Modeling and Particle Tracking through Longitudinal Gradient Bending Magnets\par}
	\vspace{1.5cm}
	{\Large Noah Bittermann\par}
	\vspace{0.5cm}
	supervised by\par
	Michael Ehrlichman 
	{ \\Summer 2015 \par}

\vspace{2cm}

\abstract{This report documents the development of a versatile model for longitudinal gradient bending magnets (LGB's) and its implementation in particle tracking simulations.  The model presented below may be used to represent an arbitrary magnetic field profile, and was successfully implemented to represent two different LGB's.  After building a symplectic integrator for our model, we were able to perform particle tracking studies.  By producing phase space plots which depicted stable trajectories, we were able to demonstrate that we had in fact correctly implemented our symplectic integration scheme.  In addition, we produced stack representations for the bending component of the magnetic field for each LGB and made preliminary comparisons between lattices implementing this model and our own.}

\end{titlepage}

\tableofcontents

\newpage

\section{Introduction}
In order to meet the needs of users in fields ranging from biophysics to material science, it is necessary to develop a new generation of synchrotron light sources.  An ideal light source provides a diffraction-limited photon beam, requiring an emittance, imprecisely speaking the spread of a particle beam in phase space,  on the order of 1 pm$\cdot$rad.  A myriad of storage ring magnet component configurations, called lattices,  exist which are designed with ultra-low emittance \cite{Intropaper}.  Reference \cite{compact} outlines the design of such a lattice in the context of the pending Swiss Light Source (SLS) upgrade, SLS2, using longitudinal gradient bending magnets.  In particular, the authors calculate the ideal bending component of a magnetic field (that is, the $B_y$ component) which minimizes emittance.  Presently, in simulating the behavior of an ultra-low emittance lattice, the LGB bending field component is modeled by a series of homogeneous field bend magnets (otherwise called a \textit{stack of S-bends} or just a \textit{stack}).  However, any physical magnet used to generate the desired longitudinally varying bend field profile will inevitably have a $B_x$ component and a $B_z$ component as well, which the stack does not properly model.  The purpose of this project was to develop a versatile model (from here on referred to as the \textit{map model}) to accurately represent all three components of an arbitrary physical LGB magnet, and to compare the new model to the stack representations by means of charged particle tracking simulations. 

\vspace{1cm}

\section{Developing a Versatile Model}

\subsection{Choosing a Class of Model}
Figures 1 and 2 depict CAD models for two different physical LGB prototypes along with their complete magnetic field profiles and isolated on-axis fields.  It was our task to accurately model the field throughout the entire volume of the beam pipe, taken as a $1$ cm radius cylinder along the z-axis.  There are a myriad of ways to model a vector valued function.  The simplest method might be to plot each component, assign it a functional form by eye, and adjust the parameters of that functional form until the best fit is found.  For example, looking to Figure 1(b), it might make sense to model the $z$ dependence of the $B_y$ component with a Gaussian function.  While this model may be adequate for this particular magnetic field map, it might be entirely inappropriate for another.  For instance, consider the $z$ dependence of the $B_y$ component in Figure 2(b).  In this case, using a Gaussian functional form would not be satisfactory.  Thus, a simple functional form does not have the versatility to model the wide variety of potential LGB magnetic fields.  


\newgeometry{left=3cm,bottom=0.1cm}
\begin{figure}[!htbp]

\centering
\caption{CAD depiction (a), complete magnetic field data (b), and on-axis magnetic field data (c) for the C-shaped LGB.  The orange rings in the CAD drawing represent superconducting coils.  For a given subplot in (b), each colored line represents a different fixed position coordinate pair.  The only true asymmetry is in the $B_y$ vs. $z$ plot.  Any others arise from the fact that the lines corresponding to some fixed position coordinate pairs are not plotted so as to not obscure the general shape of the field.}

\subfloat[C-Shaped LGB CAD Model]{
\includegraphics[scale = 0.4]{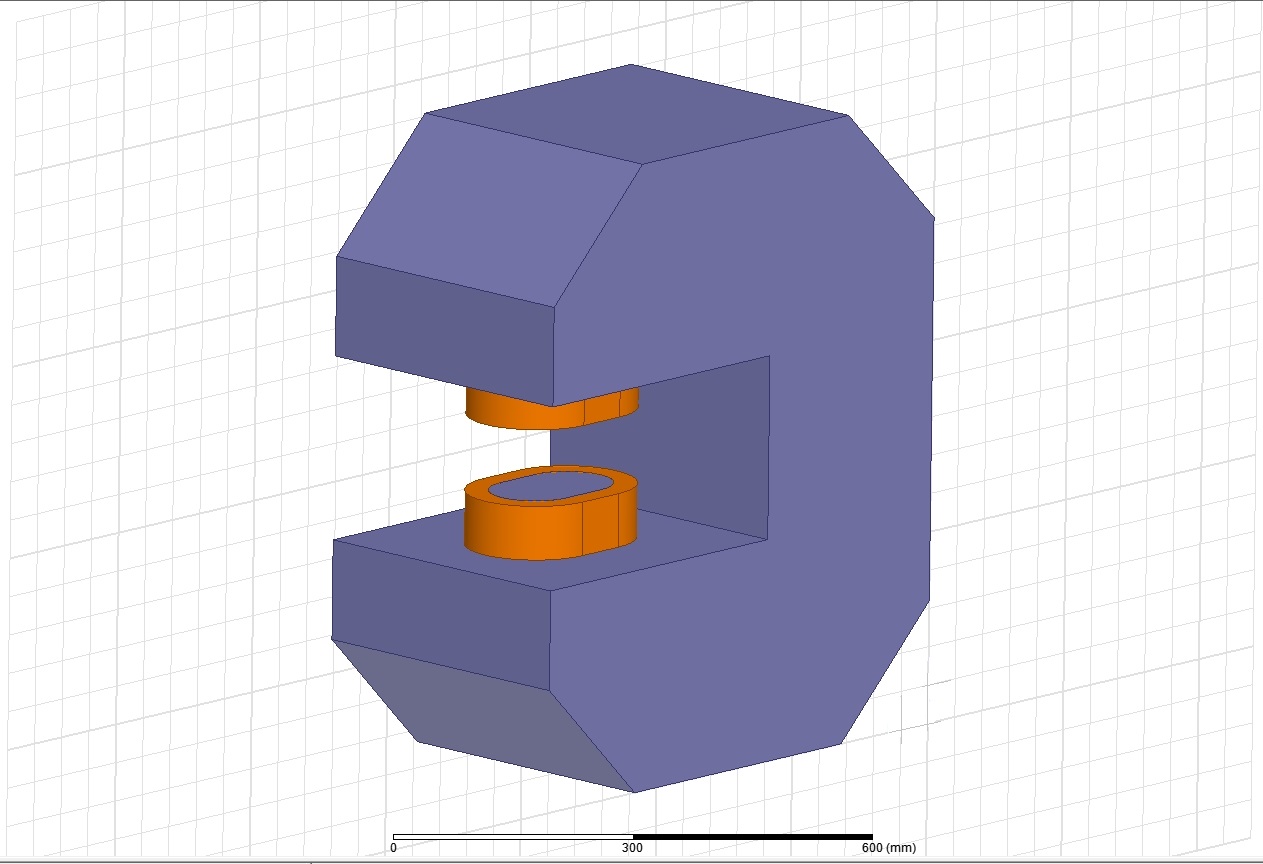}
}
\newline
\subfloat[C-Shaped LGB Magnetic Field Map]{
 \makebox[\textwidth]{\includegraphics[width=7.5in]{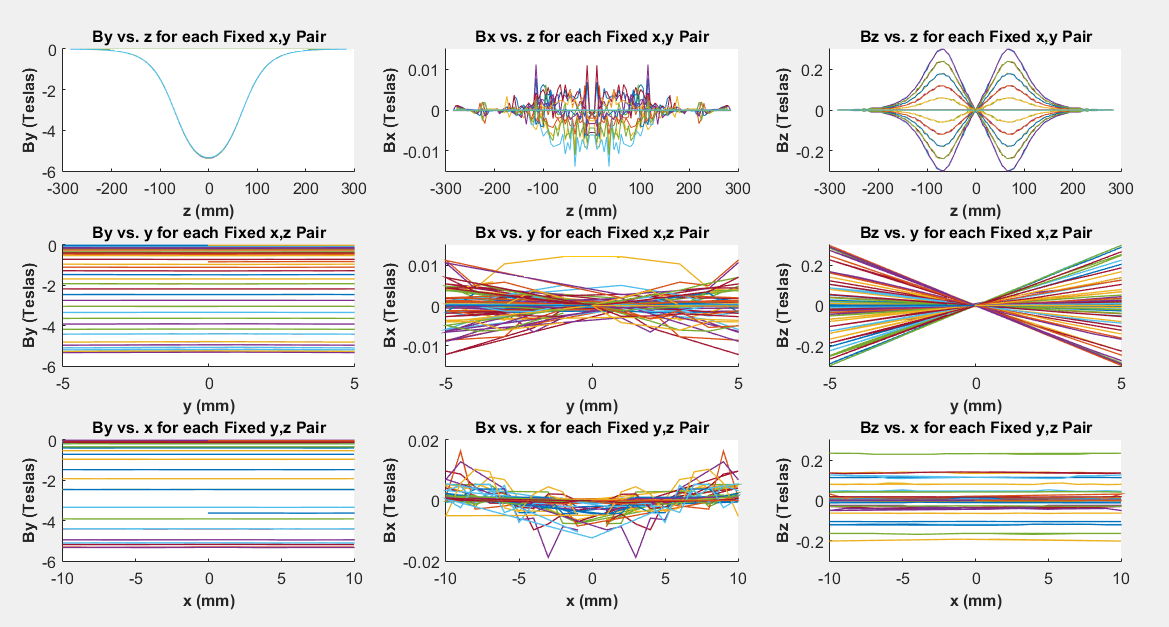}} 
}
\end{figure}

\newpage
\restoregeometry

\begin{figure}[!htbp]
\centering
\ContinuedFloat
\subfloat[C-Shaped LGB on-Axis Magnetic Field]{
 \makebox[\textwidth]{\includegraphics[width=7.5in]{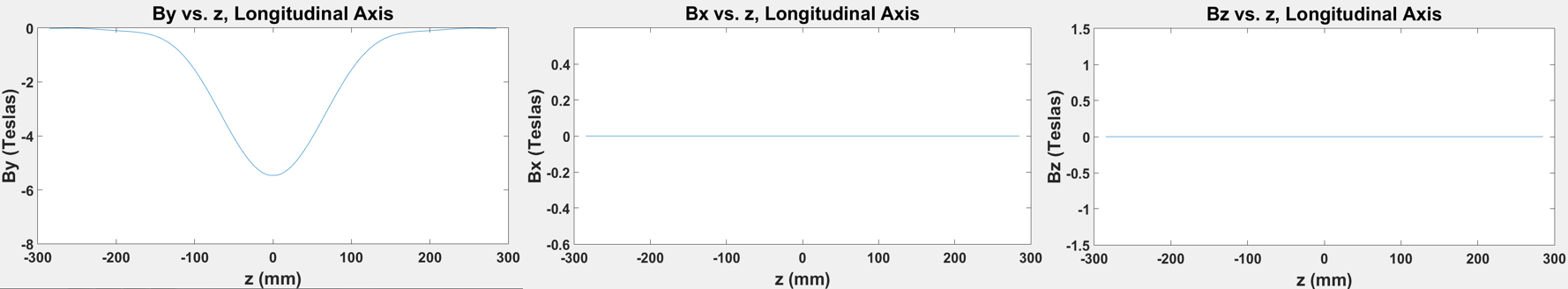}} 
}

\vspace{3cm}
\end{figure}

\setcounter{figure}{1}    
\begin{figure}[H]
\centering
\caption{CAD depiction (a), complete magnetic field data (b), and on-axis magnetic field data (c) for the canted solenoid LGB.  The red ribbons in the CAD drawing represent wire bunches.  For a given subplot in (b), each colored line represents a different fixed position coordinate pair.   The only true asymmetry is in the $B_y$ vs. $z$ plot.  Any others arise from the fact that the lines corresponding to some fixed position coordinate pairs are not plotted so as to not obscure the general shape of the field.}
\subfloat[Canted Solenoid LGB CAD Model]{
  \includegraphics[scale = 1.1]{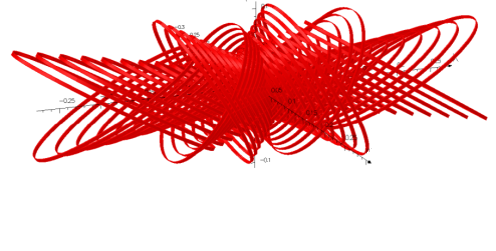}
}
\end{figure}

\newpage
\begin{figure}[H]
\centering
\ContinuedFloat
\subfloat[Canted Solenoid LGB Magnetic Field Map]{
 \makebox[\textwidth]{\includegraphics[width=7.5in]{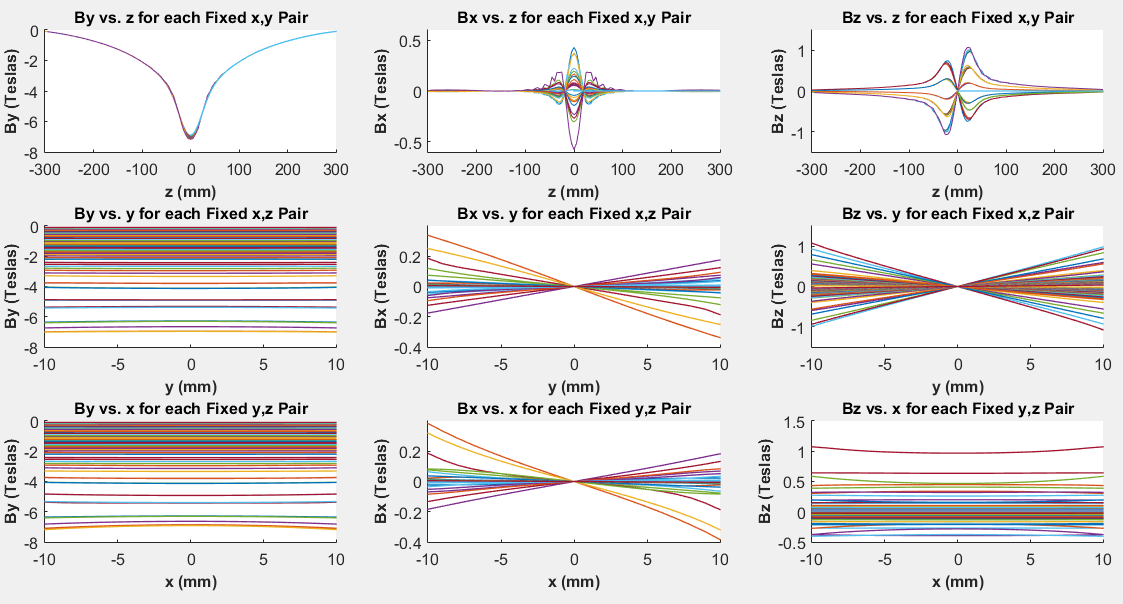}} 
}
\end{figure}

\begin{figure}[H]
\centering
\ContinuedFloat
\subfloat[Canted Solenoid LGB on-Axis Magnetic Field]{
 \makebox[\textwidth]{\includegraphics[width=7.5in]{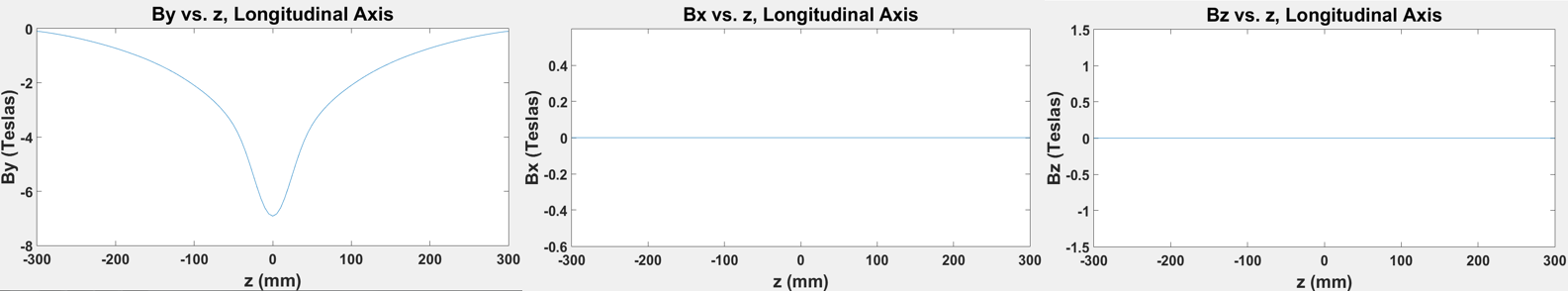}} 
}

\end{figure}

\newpage

At the sacrifice of simplicity, there does exist a class of model with the versatility to represent any magnetic field to arbitrary precision.  Instead of using a functional  form, we may fit the magnetic field data with a series of basis functions, just as one might model a wave phenomenon with a Fourier Series.  Indeed, this provides maximal versatility, since any well-behaved function can be written as an infinite sum of suitable basis functions, so long as the set of basis functions is complete in the mathematical sense.  Of course, one must terminate the series after a finite number of terms, but by including a larger and larger number of them, arbitrary precision may be achieved.  This is the course taken in reference \cite{wiggler} to model wiggler and undulator magnets.  \\

\subsection{Deriving a Suitable Set of Basis Functions}
Now that the class of model has been chosen, the problem becomes finding a proper set of basis functions.  In particular, we seek a collection of basis functions which converges to the magnetic field data quickly, and thus requires relatively few terms to achieve high precision.  With slight modifications, we employed the set of basis functions given in \cite{wiggler}.  Since the authors did not include a derivation of this basis, we provide one below.

\vspace{0.25cm}
Separable partial differential equations (PDE's) yield an ample source of candidate basis functions.  The most natural way to proceed is to solve the potential formulation of Maxwell's equations, find a set of basis functions for the vector potential, and take the curl to acquire a set for the magnetic field.  We begin with Ampere's law, expressed in terms of the magnetic vector potential $\mathbf{A}$:

\vspace{4mm}
\hspace{5cm}$\nabla(\nabla \cdot \mathbf{A}) - \nabla^2\mathbf{A} = \mu_0\mathbf{J}$,\\

\noindent in which $\mathbf{J}$ is the current density.  For the sake of computational simplicity, we choose the Coulomb gauge with the modification that $A_x = 0$.  Next we note that in our region of interest, the current density is zero.  Thus we are left with Laplace's equation:
\begin{equation}
\nabla^2\mathbf{A} = 0.
\end{equation}

The next step is to choose a coordinate system in which to solve equation (1).  Choosing different coordinate systems yields different functional bases whose corresponding series might converge with different speeds.  For similar modeling projects, it may be worthwhile to explore these different bases.  However, for our purposes it was sufficient to express equation (1) in simple Cartesian coordinates and employ the resultant basis.  As usual in solving separable PDE's, we write $A_y$ and $A_z$ as a product of three single-variable functions: $A_x(x,y,z) = H(x)I(y)J(z)$ and $A_z(x,y,z) = K(x)L(y)M(z)$.  Substituting these expressions into equation (1) and introducing separation constants, we have that \\

\begin{subequations}
\begin{align}
\frac{H''(x)}{H(x)} = -\lambda_x,  \frac{J''(z)}{J(z)} = \lambda_z,  \frac{I''(y)}{I(y)} = \lambda_x - \lambda_z = \lambda_y \\
\vspace{4mm}
\frac{K''(x)}{K(x)} = -\lambda_x,  \frac{M''(z)}{M(z)} = \lambda_z,  \frac{L''(y)}{L(y)} = \lambda_x - \lambda_z = \lambda_y      
\end{align}
\end{subequations}
\vspace{4mm}

\noindent Note that corresponding separation constants are the same for $A_y$ and $A_z$.  After deriving the full form of $A_y$ and $A_z$, taking the curl to calculate $\mathbf{B}$, and taking the divergence and curl of $\mathbf{B}$, one can see that this must be the case in order to satisfy Maxwell's equations.  In many instances of solving separable PDE's, one might be able to appeal to boundary conditions to determine that $\lambda_x$ and $\lambda_z$ are constrained to have a certain sign.  However, for the sake of remaining general, we are not able to impose boundary conditions.  Indeed, our model must be able to represent the fields of arbitrary physical magnets like those in Figures 1 and 2, which looking to the CAD models, have completely different boundary conditions.

\vspace{0.25cm}
If we were to stop here and write down every possible solution to our PDE, our model would be far too complicated to be useful.  We have that $\lambda_x$ and $\lambda_z$ may each take on three different signs (zero being a sign).  For each possible sign, $\lambda_x$ and $\lambda_z$ yield two linearly independent solutions to whichever ordinary differential equation (ODE) they are involved in.  Considering only these cases, there are already 24 different possible forms which the vector potential basis functions may take on.  That number becomes even larger when we account for the fact that in addition to the signs of $\lambda_x$ and $\lambda_z$, the sign of $\lambda_y$ also depends on their relative magnitudes.  Fortunately, with further insight and by adding a relatively small number of extra parameters we may vastly simplify the model.  First, since sine's and cosine's yield a complete basis for scalar valued functions, without loss of generality we may assert that $\lambda_z \equiv -k_z^2 < 0$, which forces $J(z)$ and $M(z)$ to be (co)sinusoidal.  Thus, we are expressing the z dependence of $\mathbf{A}$ (and thus $\mathbf{B}$) as a sum of sine's and cosine's.  Though we may have imposed this restriction on $\lambda_x$ or $\lambda_y$ it is useful to choose $\lambda_z$ for LGB fields, since qualitatively the dependence of  $\mathbf{B}$ on z looks (co)sinusoidal (see Figures 1(b) and 2(b)). 

\vspace{0.25cm}
Through an additional simplification, we may combine the cases in which $\lambda_x = 0$ and $\lambda_y = 0$ with the cases in which $\lambda_x< 0$ and $\lambda_y > 0$.  In the cases in which separation constants are zero, the ODE's in equation (2) yield solutions of the form $ax + b$.  First, we need not consider $b$, since its presence does not affect the curl of $\mathbf{A}$.  Furthermore, using the approximation that $\sinh(ku) \approx ku$ for small $k$, we may effectively consider linear solutions to the ODE's as special case of hyperbolic trigonometric solutions.  This a valid simplification because if we ever require such a linear solution, in the process of fitting the model to our data we may achieve arbitrary accuracy by choosing an arbitrarily small value for $k$.  In this instance, only being arbitrarily accurate rather than exact is allowable, since we will be limited later by the fact that we cannot fit a truly infinite series to our data, and must settle for arbitrarily many terms.  Table 1 shows the cumulation of our present simplifications.\\ \\ \\

\begin{table}[h!]
\centering
\caption{This table depicts all possible combinations of signs for $\lambda_x$ and $\lambda_z$ and the solutions to the ODE's in equations (2a) and (2b).  We do not include a column characterizing $\lambda_z$ since it is always the case that $\lambda_z \equiv -k_z^2 < 0$.  ``Trig'' and ``Hyper'' indicate that the linearly independent solutions to the ODE's in equation (2) are either of the form $\sin(k_uu)$, $\cos(k_uu)$, or $\sinh(k_uu)$, $\cosh(k_uu)$, in which $u = x, y, z$.  ``Form" is an arbitrary label used for referring to certain rows with the same relationship between $k_x$, $k_y$, and $k_z$.   This table should be read ``If conditions in columns 1 and 2, then columns 3-7."}

\resizebox{\textwidth}{!}{
\label{tab: Table 1}
\begin{tabular}{|c|c|c|c|c|c|c|}
\hline
 \rule{0pt}{3ex}
 $\lambda_x$                                  &  $k_x$, $k_z$ Relation &  $\lambda_y$                                                           &  $J(z)$ \& $M(z)$                       &  $H(x)$ \& $K(x)$                          &  $I(y)$ \& $L(y)$                            &  Form \\ \hline \rule{0pt}{3ex}  
 $\lambda_x \equiv k_x^2>0$        &  $k_x>0$                       &  $\lambda_y \equiv k_y^2 = k_x^2 + k_z^2 >0$      &  Trig   &  Trig     &  Hyper  &  1       \\ \hline \rule{0pt}{3ex}
 $\lambda_x \equiv k_x^2>0$        &  $k_x<0$                       &  $\lambda_y \equiv k_y^2 = k_x^2 + k_z^2 >0$      &  Trig   &  Trig      &  Hyper  &  1       \\ \hline \rule{0pt}{3ex}
 $\lambda_x \equiv -k_x^2<0$       &  $-|k_z| \leq k_x \leq 0$   &  $\lambda_y \equiv k_y^2 = -k_x^2 + k_z^2 >0$    & Trig      &  Hyper&  Hyper  &  2       \\ \hline \rule{0pt}{3ex}
 $\lambda_x \equiv -k_x^2<0$       &  $k_x<-|k_z|$                 &  $\lambda_y \equiv -k_y^2 = -k_x^2 + k_z^2 <0$   &  Trig       &  Hyper&  Trig     &  3       \\ \hline     

\end{tabular}}
\end{table}

\vspace{1.5cm}

With $2^3 = 8$ different combinations per row, we have reduced the total number of possible forms to 32.  To carry out our next simplification, we make two observations.  First, $\sin$ and $\cos$ differ only by a phase shift.  Second, $\sinh$ can be written as a linear combination of phase-shifted $\cosh$'s.  Likewise, $\cosh$ may be expressed in an analogous manner with $\sinh$'s.  Thus, by introducing phase shifts $\phi_x$, $\phi_y$, and $\phi_z$ to the functions in columns 4-6 of Table 1, we may reduce the number of possible forms.  As with the separation constants, in order to satisfy Maxwell's equations, the triplet of phase shifts for $A_y$ must be the same as that for $A_z$.  Furthermore, it does not matter which function in columns 4-6 of Table 1,  e.g. $\sin$ or $\cos$ in the case of ``Trig", we endow with the phase shift.  However, anticipating the qualitative symmetry of the LGB magnetic fields to which we are fitting, it is best to make the choices presented in Table 2, where we now distinguish between $A_y$ and $A_z$.  While we remain general for now and include the phase shifts, with the choices of functions in Table 2 it is possible to fix the phase shifts to be zero due to the aforementioned symmetry.  In addition, adding phase shifts allows us to consider row 2 in Table 1 as a special case of row 1.  Finally, since the $k$'s are actually used as arguments for the functions in columns 4-6 of Table 1 rather than the $\lambda$'s, we use them to characterize Table 2.  With these modifications, though we have introduced three new parameters, we have reduced the number of possible forms for $\mathbf{A}$ to 3.\\ \\ \\

\begin{table}[h!]
\centering
\caption{This table depicts all possible order relations between $k_x$, $k_y$, and $k_z$ and modifications to the functions in Table 1 by adding phase shifts.  The top portion represents $A_y$ while the bottom represents $A_z$.  This table should be read ``If conditions in columns 1-2, then columns 3-6."}

\vspace{5mm}

\resizebox{\textwidth}{!}{
\label{tab: Table 2}
\begin{tabular}{|c|c|c|c|c|c|c|c|}
\hline
\multicolumn{6}{|c|}{$A_y$} \\ \hline
 \rule{0pt}{3ex}
$k_x$, $k_y$ Relation  &  $k_z$                                  &            $J(z)$ \& $M(z)$       &              $H(x)$ \& $K(x)$       &           $I(y)$ \& $L(y)$          &  Form      \\ \hline \rule{0pt}{3ex}  
$k_x>0$                       &  $k_y^2 = k_x^2 + k_z^2$    &        $\sin(k_zz + \phi_z)$      &        $\sin(k_x x + \phi_x)$      &       $\sinh(k_yy + \phi_y)$      &  1            \\ \hline \rule{0pt}{3ex}
$-|k_z| \leq k_x \leq 0$             &  $k_y^2 = k_z^2 - k_x^2$    &        $\sin(k_zz + \phi_z)$      &        $\sinh(k_x x + \phi_x)$    &       $\sinh(k_yy + \phi_y)$      &  2            \\ \hline \rule{0pt}{3ex}
$k_x<-|k_s|$                 &  $k_y^2 = k_x^2 - k_z^2 $   &        $\sin(k_zz + \phi_z)$      &        $\sinh(k_x x + \phi_x)$    &       $\sin(k_yy +   \phi_y)$      &  3            \\ \hline     

\multicolumn{6}{|c|}{$A_z$} \\ \hline
 \rule{0pt}{3ex}
$k_x$, $k_y$ Relation  &  $k_z$                                  &            $J(z)$ \& $M(z)$       &              $H(x)$ \& $K(x)$       &           $I(y)$ \& $L(y)$             &  Form      \\ \hline \rule{0pt}{3ex}  
$k_x>0$                       &  $k_y^2 = k_x^2 + k_z^2$    &        $\cos(k_zz + \phi_z)$      &        $\sin(k_x x + \phi_x)$      &       $\cosh(k_yy + \phi_y)$      &  1            \\ \hline \rule{0pt}{3ex}
$-|k_z| \leq k_x \leq 0$             &  $k_y^2 = k_z^2 - k_x^2$    &        $\cos(k_zz + \phi_z)$      &        $\sinh(k_x x + \phi_x)$    &       $\cosh(k_yy + \phi_y)$      &  2            \\ \hline \rule{0pt}{3ex}
$k_x<-|k_s|$                 &  $k_y^2 = k_x^2 - k_z^2$   &        $\cos(k_zz + \phi_z)$      &        $\sinh(k_x x + \phi_x)$    &       $\cos(k_yy +   \phi_y)$      &  3            \\ \hline     

\end{tabular}}
\end{table}

\newpage
We are almost able to put forth a set of basis functions for $\mathbf{A}$.  However, we must first multiply $A_y$ and $A_z$ by an appropriate scale factor, determined by the fact that $\mathbf{B}$ must have zero curl and divergence.  A set of basis functions for $\mathbf{A}$ is presented below, with the proper scale factors. 
\vspace{5mm}

$$
\mathbf{A} =
\begin{cases}
\text{Form } 1 & \text{if }  k_x>0       \\
\text{Form } 2 & \text{if }  -|k_z| \leq k_x \leq 0 \\
\text{Form } 3 & \text{if }  k_x<-|k_z|  
\end{cases}
$$

\begin{table}[h!]

\centering
\begin{tabular}{l}
                                        Form 1                                      \\
\hline                                              
$A_x = 0$                                                                     \\                                                    
$A_y = -\frac{k_z}{k_x k_y}\sin(k_x x + \phi_x)\sinh(k_yy + \phi_y)\sin(k_zz + \phi_z)$  \\
$A_z = -\frac{1}{k_x}\sin(k_x x + \phi_x)\cosh(k_yy + \phi_y)\cos(k_zz + \phi_z)$ \\           
\hspace{1cm} with $k_y^2 = k_x^2 + k_z^2$                                                                                   \\ \\
			        Form 2 \\
\hline
$A_x = 0$     \\
$A_y = -\frac{k_z}{k_x k_y}\sinh(k_x x + \phi_x)\sinh(k_yy + \phi_y)\sin(k_zz + \phi_z)$ \\
$A_z = -\frac{1}{k_x}\sinh(k_x x + \phi_x)\cosh(k_yy + \phi_y)\cos(k_zz + \phi_z)$ \\
\hspace{1cm} with $k_y^2 = k_z^2 - k_x^2$ \\ \\
			        Form 3 \\
\hline
$A_x = 0$ \\
$A_y =-\frac{k_z}{k_x}\sinh(k_x x + \phi_x)\sin(k_yy + \phi_y)\sin(k_zz + \phi_z)$ \\
$A_z = -\frac{1}{k_x}\sinh(k_x x + \phi_x)\cos(k_yy + \phi_y)\cos(k_zz + \phi_z)$  \\
\hspace{1cm} with $k_y^2 = k_x^2 - k_z^2$\\

\end{tabular}
\end{table}

Finally, taking the curl of $\mathbf{A}$ gives us $\mathbf{B}$, which is presented below.
\newpage

\begin{equation}
\mathbf{B} =
\begin{cases}
\text{Form } 1 & \text{if }  k_x>0       \\
\text{Form } 2 & \text{if }  -|k_z| \leq k_x \leq 0 \\
\text{Form } 3 & \text{if }  k_x<-|k_z|    
\end{cases}
\end{equation}

\begin{table}[h!]

\centering
\begin{tabular}{l}
                                        Form 1   \\
\hline                                                                                     
$B_x = -\frac{k_x}{k_y}\sin(k_x x + \phi_x)\sinh(k_yy + \phi_y)\cos(k_zz + \phi_z)$     \\ 
$B_y = \cos(k_x x + \phi_x)\cosh(k_yy + \phi_y)\cos(k_zz + \phi_z)$                              \\
$B_z = -\frac{k_z}{k_y}\cos(k_x x + \phi_x)\sinh(k_yy + \phi_y)\sin(k_zz + \phi_z)$       \\
\hspace{1cm}  with $k_y^2 = k_x^2 + k_z^2$                                                                   \\ \\ 
Form 2 \\
\hline
$B_x = \frac{k_x}{k_y}\sinh(k_x x + \phi_x)\sinh(k_yy + \phi_y)\cos(k_zz + \phi_z)$   \\  
$B_y = \cosh(k_x x + \phi_x)\cosh(k_yy + \phi_y)\cos(k_zz + \phi_z)                      $\\
$B_z = -\frac{k_z}{k_y}\cosh(k_x x + \phi_x)\cosh(k_yy + \phi_y)\cos(k_zz + \phi_z)$\\
\hspace{1cm}  with $k_y^2 = k_z^2 - k_x^2$ \\\\
Form 3 \\
\hline
$B_x = \frac{k_x}{k_y}\sinh(k_x x + \phi_x)\sin(k_yy + \phi_y)\cos(k_zz + \phi_z)$ \\
$B_y =\cosh(k_x x + \phi_x)\cos(k_yy + \phi_y)\cos(k_zz + \phi_z)$ \\
$B_z = -\frac{k_z}{k_y}\cosh(k_x x + \phi_x)\sin(k_yy + \phi_y)\sin(k_zz + \phi_z)$ \\
\hspace{1cm} with $k_y^2 = k_x^2 - k_z^2$

\end{tabular}
\end{table}

\vspace{1cm}
We include the caveat that if $k_x$ and $k_z$ are both zero, then $\frac{k_x}{k_y}$ and $\frac{k_z}{k_y}$ should be replaced with $1$ so that our basis functions are continuous in parameter space.  With this, we have obtained a basis which is suitable for modeling any magnetic field profile.  This basis may be further simplified for different types of magnets.  For example, given the symmetry of wiggler magnets, one may impose $\phi_x$ and $\phi_y$ to be zero, as is done in \cite{wiggler}.  For modeling LGB's we make the following simplifications.  First, looking to Figures 1(b) and 2(b), it is typical for $B_x$ vs. $x$, $y$, and $z$ and $B_z$ vs. $x$, $y$, and $z$ to be centered on and symmetric about their horizontal axes.  This is not the case for $B_y$.  To capture this behavior, we require a constant offset in $B_y$, just as one might require a constant offset in modeling a scalar function with a Fourier series.  While a constant offset is already built into the aforementioned basis, it was difficult for our optimization program to properly converge to a reasonable value without isolating it.  We may isolate the necessary constant offset in $B_y$ in a simple manner: Impose that for the first term in the series $k_x = k_y = \phi_x = \phi_y = \phi_z = 0$.  Since $k_x = 0$, we use Form 2 for this term.  Furthermore our impositions force $k_y = 0$, so we have that $B_y = 1$, while $B_x = B_z = 0$.  After this is done, given the symmetry of LGB fields, the phase shifts are no longer needed and can be set to zero for all other terms.  The cumulation of our derivation and simplifications is that we obtain the follow series to model our magnetic field data:\\
\begin{equation}
\mathbf{B}_{fit}(x,y,z) = 
\begin{bmatrix}
0\\b_y\\0
\end{bmatrix}
 + \displaystyle\sum_{i = 2}^{N} C_i \mathbf{B}_i (x,y,z;k_{xi},k_{zi}),
\end{equation}

\noindent in which $\mathbf{B}_i$ is one of the basis functions in equation (3) and $b_y$, $C_i$, $k_{xi}$, and $k_{zi}$ are our fit parameters. 

\vspace{1.5cm}
\section{Fitting the Model to Data}
In this project, we were given the task of fitting the magnetic field data of two different LGB prototypes.  The LGB  prototypes were developed by a group specializing in magnet design using a CAD program.  CAD models for each component and their respective magnetic field maps are presented in Figures 1 and 2.  Because the $B_x$ component for the C-shaped LGB is so small compared to the other components of the field ($\sim$10 times smaller than the $B_z$ component and $\sim$100 times smaller than $B_y$ component) and so disorderly, we took it to be noise, and set it to be zero before fitting to the complete field.  On the other hand, the $B_x$ component for the canted solenoid LGB is not small enough to justify setting it to zero before fitting.

\vspace{0.25cm}
To perform the fitting, we coded equation (4) into MATLAB and employed one of its built-in optimizers.  We had the option of using a trust-region-reflective algorithm or a Levenberg-Marquardt algorithm.  A trust-region algorithm first approximates the function one wishes to optimize with a simpler function.  It then proceeds to optimize this new function.  In particular, the algorithm requires the function being optimized to have more components than fit parameters \cite{trust}.  However, our model function is underdetermined--it has only three components, but $3N-2$ parameters for $N$ terms.  Thus the trust-region algorithm was unviable, and we used a Levenberg-Marquardt algorithm instead.  This algorithm is a local optimizer, meaning that in order to converge to a set of optimal fit parameters, the initial fit parameters fed into the algorithm must be in the catchment basin of a reasonable local minimum.  This places a large significance on the initial guess for the fit parameters.  For instance, if one were to try and fit to the canted solenoid field using an arbitrary initial guess of $C_i = k_{xi} = k_{yi} = b_y = 1$ for all $i$, then the optimizer would not converge.  Looking at the field data, it is difficult to determine what a suitable initial guess might be.  We got around this by splitting the optimization into parts.  We first fit the model to just one component of the magnetic field, then used this fit as the initial guess for the fit of the entire field.  In the case of both LGB's, by ignoring the $B_y$ and $B_z$ components we were able to adequately fit to the $B_x$ component of the field with the arbitrary initial parameters guess of  $C_i = k_{xi} = k_{yi} = b_y = 1$.  We were then able to fit to the entire field using the result of our $B_x$ fit as an initial guess.  Figure 3 shows plots of the fits for both the C-shaped and canted solenoid LGB's.  Figure 4 plots the residuals for each component of both LGB's.  \\ \\

\vspace{1cm}

\setcounter{figure}{2}    
\begin{figure}[H]
\centering
\caption{Fit plots for the C-shaped (a) and canted solenoid (b) LGB's.  As in the plots of the real data, differently colored lines correspond to different fixed coordinate pairs.  For the C-shaped magnet, $B_x$ was set to zero before fitting.  Again, lines corresponding to some fixed coordinate pairs are not plotted so as to not obscure the general shape of the field.  The quality of these fits is quantified in Table 3.}
\subfloat[C-Shaped LGB Fit-Field]{
 \makebox[\textwidth]{\includegraphics[width=7.5in]{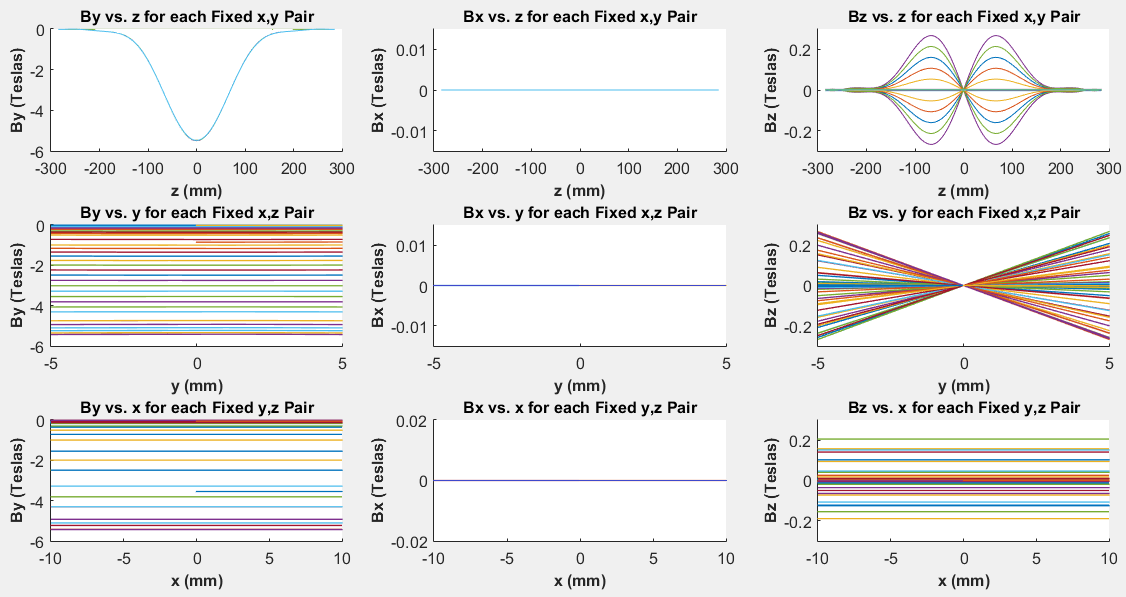}} 
}
\end{figure}

\begin{figure}[H]
\centering
\ContinuedFloat
\subfloat[Canted Solenoid LGB Fit-Field]{
 \makebox[\textwidth]{\includegraphics[width=7.5in]{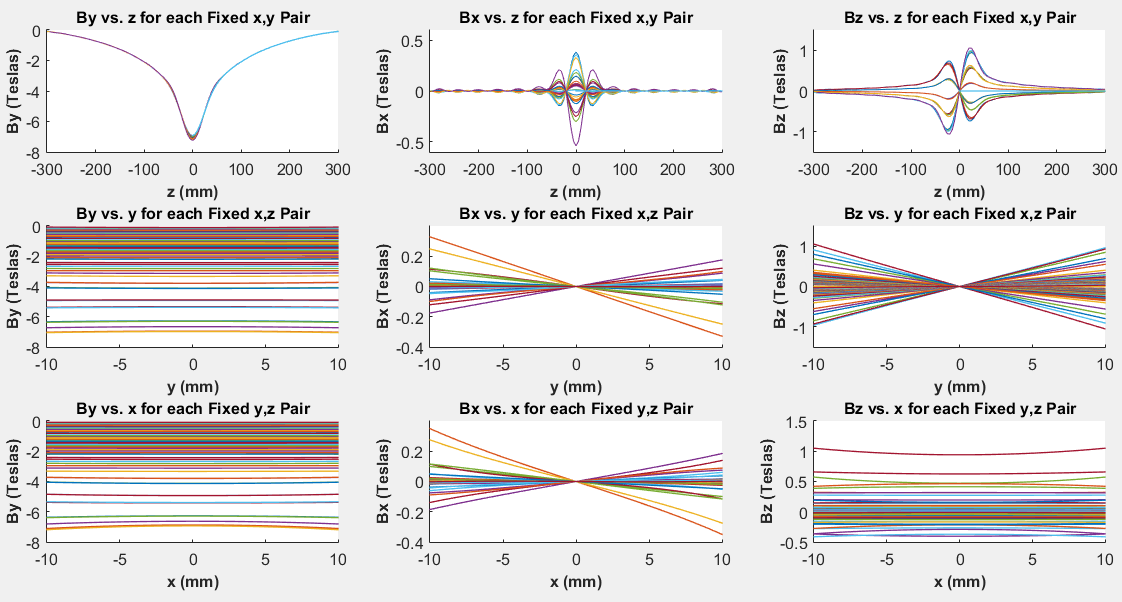}} 
}

\end{figure}

\setcounter{figure}{3}    
\begin{figure}[H]
\caption{Residual plots for the C-shaped (a) and canted solenoid (b) LGB's.}
\centering
\subfloat[C-Shaped LGB Residual Plots]{
 \makebox[\textwidth]{\includegraphics[width=7.5in]{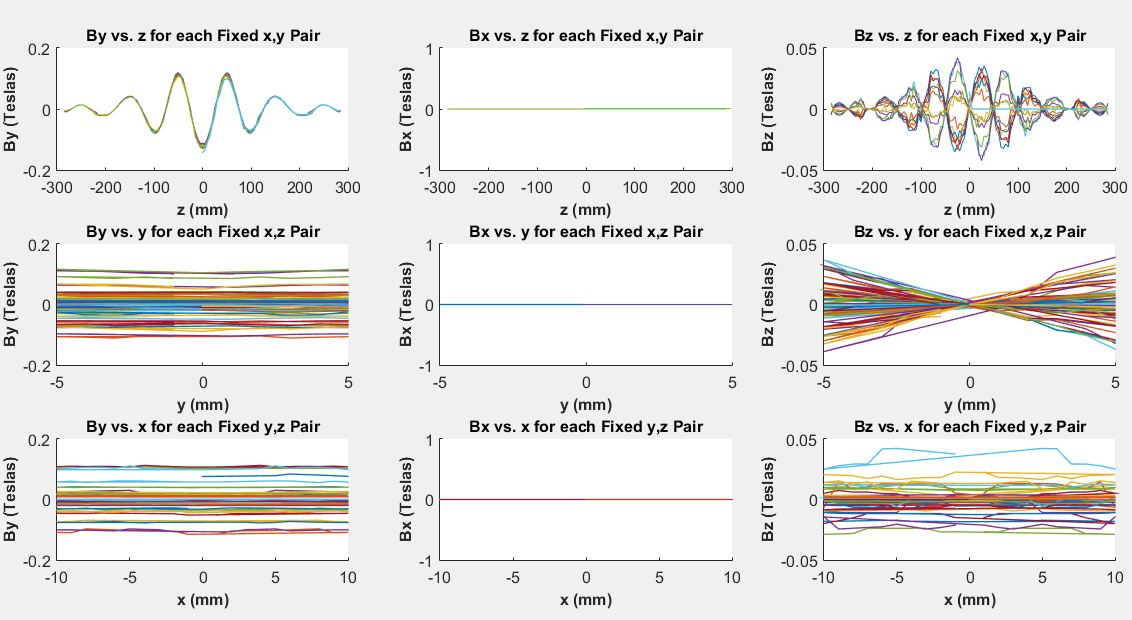}} 
}
\end{figure}
\begin{figure}[H]
\centering
\ContinuedFloat
\subfloat[Canted Solenoid LGB Residual Plots]{
 \makebox[\textwidth]{\includegraphics[width=7.5in]{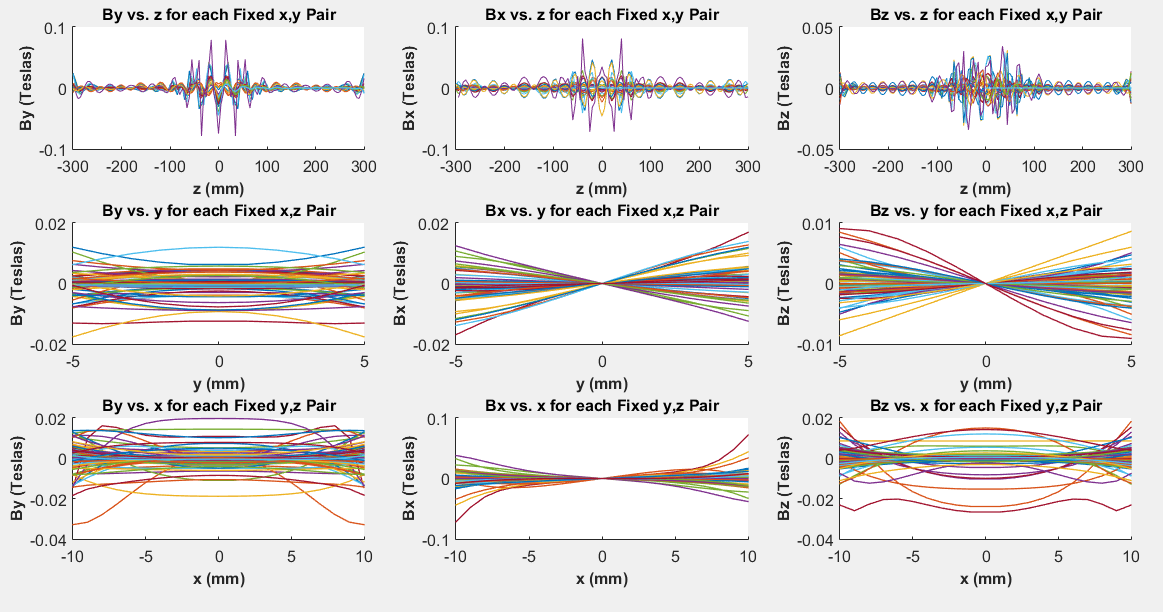}} 
}

\end{figure}

\vspace{2cm}

The quality of each fit is quantified in Table 3.  We used as many terms in the series as was necessary to drive the average residual between each component of the fit field and the true field below 10\% of the magnitude of the true field.  For instance, using 7 terms for the C-shaped LGB fit yields an average residual of $0.000$ $T$ in $B_x$, $0.036$ $T$ in $B_y$, and $0.007$ $T$ in $B_z$, each of which is less than 10\%  of the magnitude of the true $B_x$, $B_y$, and $B_z$ components.  Looking to Table 3, the maximum residual is simply the magnitude of the maximum difference between the fit field and the true field.  To calculate the average residual between the fit field and the true field, we simply summed the absolute values of the residuals for each magnetic field component, and divided by the total number of data points. \\

\begin{table}[H]   
\centering
\caption{Quantitative evaluation of fit results for the C-shaped and canted solenoid LGB's.  The constant offset $b_y$ is counted as its own term.}  
\begin{tabular}{|l|l|l|l|}

  \hline
  C-Shaped LGB--7 Terms & $B_x$ $(T)$ & $B_y$ $(T)$ & $B_z$ $(T)$ \rule{0pt}{3ex}  \\ 
  \hline
  Maximum Residual Magnitude & 0.0 & 1.4\e{-1} & 4.2\e{-1} \rule{0pt}{3ex} \\
  \hline
  Sum of Residual Magnitudes & 0.0 & 8.6\e{2} & 1.6\e{2} \rule{0pt}{3ex} \\
  \hline
  Average Residual (23,776 data points) & 0.0 & 3.6\e{-2} & 6.8\e{-3} \rule{0pt}{3ex} \\
  \hline
  Canted Solenoid LGB--9 Terms & $B_x$ $(T)$ & $B_y$ $(T)$ & $B_z$ $(T)$ \rule{0pt}{3ex}  \\

  \hline
  Maximum Residual Magnitude & 8.1\e{-2} & 7.9\e{-2} & 3.4\e{-2} \rule{0pt}{3ex}\\
  \hline
  Sum of Residual Magnitudes & 1.78\e{2} & 2.0\e{2} & 1.3\e{2} \rule{0pt}{3ex} \\
  \hline
  Average Residual (50,900 data points) & 3.0\e{-3} & 3.3\e{-3}  & 2.1\e{-3} \rule{0pt}{3ex} \\
  \hline

\end{tabular}
\end{table}

\section{Building a Symplectic Integrator}
\subsection{Symplectic Integration Overview}
In this section, we give a brief introduction to symplectic integration and explain why it is a necessary procedure in particle tracking simulations.  We begin with a fact from Hamiltonian mechanics:  Hamiltonian flows are symplectic, meaning they preserve the area associated with the following two-form:

\begin{center}
$\omega = \displaystyle\sum_{i = 1}^{3} dp_i \wedge dq_i$,
\end{center}

\noindent in which $q_i$ is a generalized coordinate and $p_i$ is its conjugate momentum \cite{twoform}.  In particular, this means that given some closed trajectory in 6-dimensional phase space, the \textit{sum} of the area of the projections onto the $x$, $p_x$ plane, the $y$, $p_y$ plane, and the $z$, $p_z$ plane is constant for all time, though the area of each individual projection may change.

\vspace{0.25cm}
As an example, consider a simple 1-dimensional  storage ring consisting of a drift space, a focusing quad, a second drift space, and a defocusing quad.  For each element Hamilton's equations may be solved analytically.  A particle which makes a single pass through this ring traces a closed trajectory in $x$, $p_x$ phase space.  As the particle continues to circulate through the ring, the area of this trajectory does not change.  In particular, the particle's trajectory in phase space does not spiral in or out.  The first plot in Figure 5(a) depicts the particle's trajectory over 1000 circulations through the ring, determined by solving Hamilton's equations analytically.

\vspace{0.5cm}

\setcounter{figure}{4}    
\begin{figure}[H]
\centering
\caption{Phase Space plots for a 1000 turns in simple ring consisting of a drift space, a focusing quad, a second drift space, and a defocusing quad.  These plots were produced by solving Hamilton's equations exactly, numerically with Euler's method, numerically with a Runga--Kutta scheme, and numerically using a symplectic integrator.}
 \makebox[\textwidth]{
\subfloat[]{
  \includegraphics[height = 4.25cm]{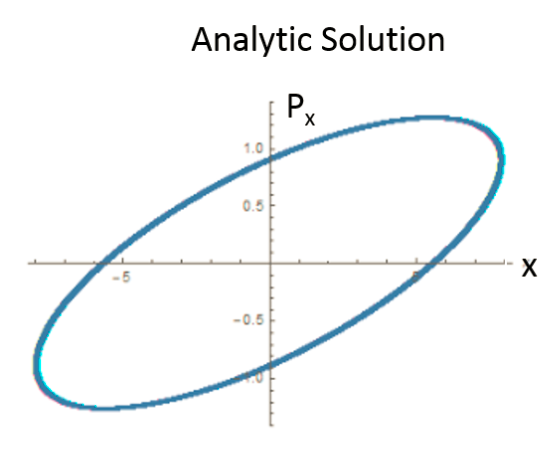}
}
\subfloat[]{
  \includegraphics[height = 4cm]{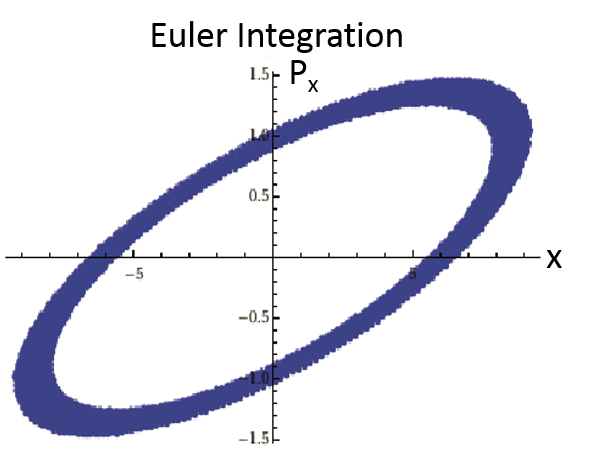}
}
\subfloat[]{
  \includegraphics[height = 4cm]{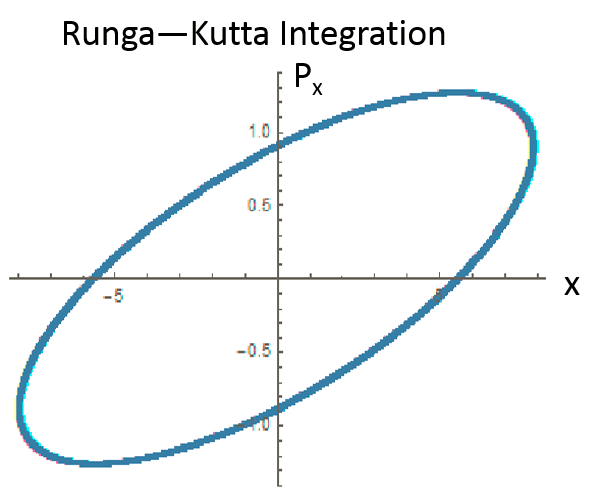}
}}
\newline
 \makebox[\textwidth]{
\subfloat[]{
  \includegraphics[height = 4cm]{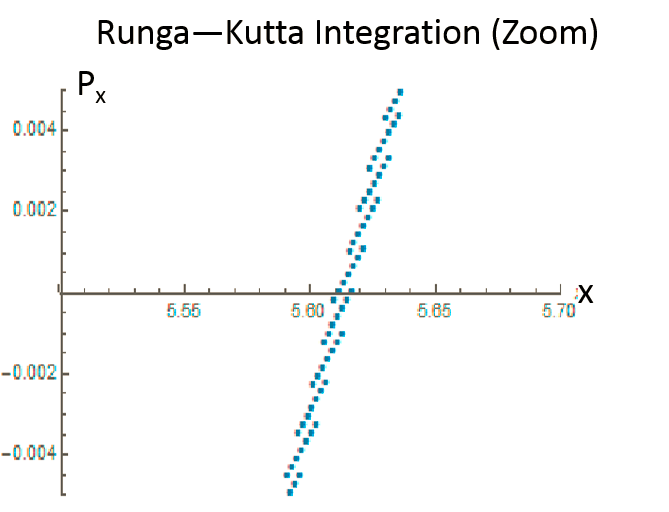}
}
\subfloat[]{
  \includegraphics[height = 4cm]{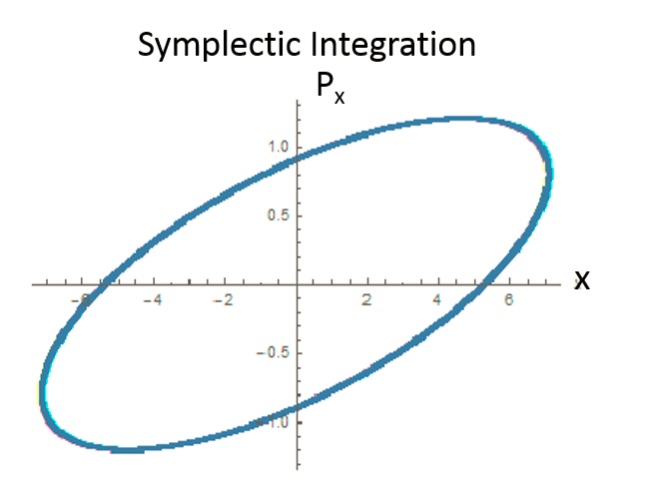}
}
\subfloat[]{
  \includegraphics[height = 4cm]{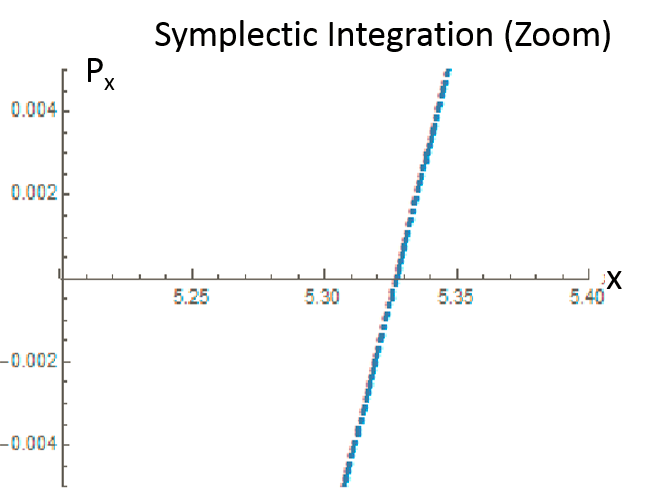}
}}

\end{figure}

\vspace{1cm}

\vspace{0.25cm}
For more complicated systems, Hamilton's equations cannot be solved analytically.  Therefore, we must solve them using numeric integration schemes.  In order for our simulations to be physical, it is essential that whatever scheme we use also preserves the area associated with the aforementioned two-form.  Such a scheme is called a symplectic integrator.  If we do not use a symplectic integrator, then our tracking simulations do not accurately depict the stability of a particle beam in a storage ring.  If the $x$ coordinate of a simulated particle grew without bound so that the particle was lost from the ring, it would be impossible to know if this were due to a numerical artifact or the design of the ring itself. 

\vspace{0.25cm}
To demonstrate the adverse effects of not using a symplectic integrator, we refer back to our simple example.  Figure 5(b) depicts the trajectory of a simulated particle in phase space for the case in which we solved Hamilton's equations numerically using Euler's method, which is not symplectic.  Evidently, the particle's trajectory spirals outward.  Since we are tracking through the same ring as before, this behavior must be an artifact of our integration scheme.  Figure 5(c) depicts the trajectory of simulated particle in which we solved Hamilton's equations numerically using a Runga-Kutta scheme.  While at first glance it may appear that this scheme is symplectic and preserves the area of the ellipse in phase space, zooming-in (Figure 5(d)) demonstrates that the trajectory does indeed spiral outward, though to a less extreme degree than in the case of Euler's method.  Figure 5(e) depicts the particle's trajectory for the case in which we solved Hamilton's equations numerically using a symplectic method.  In particular, one should note that there is no outward spiral, even as we zoom-in (Figure 5(f)).

\subsection{Deriving a Simple Symplectic Integrator}
Next, we sketch the derivation of a symplectic integrator.  What follows is a paraphrasing of \cite{Liesite} and \cite{Liepaper}.  For the sake of simplicity, we proceed in only one dimension.  First, let $\mathcal{M}$ be the map corresponding to the solution of Hamilton's equations for a Hamiltonian $H$.  We may express our final variables, $x(\Delta s)$ and $p_x(\Delta s)$, in terms of our initial variables, $x(0)$ and $p_x(0)$, for some distance $\Delta s$ through a magnet component by

\vspace{2mm}
\begin{center}
$\begin{bmatrix} x(\Delta s) \\ p_x(\Delta s) \end{bmatrix} = \mathcal{M} \begin{bmatrix} x(0) \\ p_x(0) \end{bmatrix}$.
\end{center}

\noindent Our goal is to derive a symplectic approximation to the map $\mathcal{M}$.  To do this, we first expand our final variables using a Taylor series centered around $s = 0$.  For simplicity, we do this only for $x$:

\vspace{2mm}

$$x(\Delta s) = \displaystyle\sum_{n=0}^{\infty} \frac{\Delta s^n}{n!} \frac{d^nx}{ds^n}\Bigr|_{\substack{s = 0}}.$$

\vspace{2mm}
To proceed, we put forth the following identity, whose proof is a simple application of the chain rule.  Here \{\} denotes the Poisson Brackets.

\begin{center}
$\{- \Delta s H, x\}  = -\Delta s \frac{\partial H}{\partial x}\frac{\partial x}{\partial p_x} + \Delta s \frac{\partial H}{\partial p_x}\frac{\partial x}{\partial x} =  \Delta s \frac{dp_x}{ds}\frac{\partial x}{\partial p_x} + \Delta s \frac{dx}{ds}\frac{\partial x}{\partial x} = \Delta s \frac{dx}{ds}$.
\end{center}

\vspace{2mm}
\noindent A simple mathematical induction yields

\vspace{2mm}
\begin{center}
$\{- \Delta s H, x\}^n = \{-\Delta s H, \{ -\Delta s H ,\{..., \{-\Delta s H,x\}...\}\}\} =  \Delta s^n \frac{d^nx}{ds^n}$,
\end{center}

\vspace{2mm}
\noindent and thus our Taylor expansion may be written

\vspace{2mm}
\begin{center}
 $x(\Delta s) = \displaystyle\sum_{n=0}^{\infty} \frac{\Delta s^n}{n!} \frac{d^nx}{ds^n}\Bigr|_{\substack{s = 0}} = \displaystyle\sum_{n=0}^{\infty} \frac{\{-\Delta s H,x\}^n}{n!}$.
\end{center}
\vspace{2mm}
We now define the Lie exponential operator: $\exp(:f:)g \equiv \sum_{n=0}^{\infty} \frac{\{f,g\}^n}{n!}$, so that \\

\begin{center}
$x(\Delta s) = \exp(-\Delta s:H:)x(0)$.
\end{center}

\vspace{2mm}

\noindent Therefore, since $x(\Delta s) = \mathcal{M}x(0)$, we have that 

\vspace{2mm}
\begin{equation}
\mathcal{M} = \exp(- \Delta s:H:).
\end{equation}

\noindent Though we do not prove it, we note that Lie exponential operators and their compositions are symplectic.  The rest of the derivation involves manipulating Lie exponential operators.  In particular, suppose that we have a Hamiltonian which may be written $H = H_1 + H_2$.  Invoking the Campbell-Baker-Hausdorff (CBH) theorem we may write

\begin{multline}
\mathcal{M} = \exp(-\Delta s:H_1 + H_2:) = \exp(-\frac{\Delta s}{2}: H_1:)\exp(-\Delta s :H_2:) \\
\exp(-\frac{\Delta s}{2}: H_1:) + O(\Delta s^3).
\end{multline}

This is the final form of our simple symplectic integrator--all that is left to do is evaluate the composition of Lie exponential operators, which requires a particular Hamiltonian.  It is useful to turn an exponential of a sum into a product of exponentials because it is typically easier to evaluate Lie exponential operators for $H_1$ and $H_2$ and compose them than it is to evaluate a single exponential operator for $H$.  In fact, one method of evaluating a Lie exponential operator for an arbitrary Hamiltonian is analytically solving Hamilton's equations; according to equation (5) the two are one in the same.  Thus, what we are effectively doing is decomposing the Hamiltonian into different terms, analytically solving Hamilton's equations for each term, and composing the resulting maps as in equation (6). \\   

\subsection{Modifications to Existing Bmad Symplectic Integration Software}
Before this project, there existed a Bmad symplectic integrator program specifically designed for wiggler magnets.  In particular, looking back to equation (3), the existing integrator assumed $\phi_x$ and $\phi_y$ to be zero for every term, and did not employ a constant offset as we do in our model.  Since we could simply set $\phi_z$ to be zero when we fed our fit parameters into the integrator program, the only change we had to make to the existing integrator was in accounting for the constant offset $b_y$. To demonstrate the changes we made to the Bmad symplectic integrator, we first present the Hamiltonian of a particle traveling through an LGB (and also a wiggler magnet) \cite{bmad}

\vspace{2mm}
\begin{center}
$H = H_x + H_y + H_z$,
\end{center}

\noindent with\\

\vspace{2mm}
\begin{center}
$ H_x = \frac{(p_x - a_x)^2}{2(1+p_z)}$, \hspace{0.5 cm} $H_y = \frac{(p_y - a_y)^2}{2(1+p_z)}$, \hspace{0.5 cm} $H_z = -a_z$,\\ 
\end{center}
\vspace{2mm}
\noindent in which \\
\vspace{2mm}
\hspace{3cm}$ a_x = 0, \hspace{10mm} a_y =  \frac{q}{P_0c} \int\limits_0^x B_z dx', \hspace{4mm}    a_z =  -\frac{q}{P_0c} \int\limits_0^x B_y dx' ,$\\
are the components of the vector potential, normalized by the particle's charge, $q$, the speed of light, $c$, and the reference momentum, $P_0$.  The form of our symplectic integrator is

\begin{align}
\mathcal{M} = T_{s/2} \hspace{1mm} I_{x/2} \hspace{1mm} I_{y/2} \hspace{1mm} I_z \hspace{1mm} I_{y/2} \hspace{1mm} I_{x/2} \hspace{1mm} T_{s/2} +  O(\Delta s^3)
\end{align}
\noindent in which \\
\vspace{.5mm}
\begin{center}
$T_{s/2}: s\rightarrow s+\frac{\Delta s}{2}$,\\
\end{center}
\vspace{.5mm}
\begin{center}
$I_{x/2} = \exp(:-\frac{\Delta s}{2} H_x:)$,\\
\end{center}
\vspace{.5mm}
\begin{center}
$I_{y/2} = \exp(:-\frac{\Delta s}{2} H_y:)$, \\
\end{center}
\vspace{.5mm}
\begin{center}
$I_{z} = \exp(:-\Delta s H_z:)$.
\end{center}

\noindent To evaluate $I_{x/2}$ and $I_{y/2}$, one would need to solve Hamilton's equations corresponding to the Hamiltonians $H_x$ and $H_y$.  This is difficult to do directly.  However, we may simplify matters by noting that  

\vspace{2mm}
\begin{center}
$\displaystyle \frac{(p_y - a_y)^2}{2(1+p_z)} =  \exp(:-\displaystyle\int_{0}^{y} a_y dy':)\frac{p_x^2}{2(1+p_z)}$
\end{center}
\vspace{2mm}

\noindent which may be proven by simply carrying out the definition of the Lie exponential.  Next, we apply the well-known Lie algebraic identity

\vspace{2mm}    
\begin{center}
$\exp(:A:)\exp(:B:)\exp(:-A:) = \exp(:\exp(:A:)B:)$ 
\end{center}
\vspace{2mm}

\noindent and find that 

\vspace{2mm}
\begin{center}
\begin{multline*}
I_{y/2} =  \exp(:-\frac{\Delta s}{2} H_y:) \\
=  \exp(:-\displaystyle\int_{0}^{y} a_y dy':)\exp(:-\frac{\Delta s}{2}\frac{p_x^2}{2(1+p_z)}:)\exp(:\displaystyle\int_{0}^{y} a_y dy':).
\end{multline*}
\end{center}

\noindent The argument of the middle exponential is a simple Hamiltonian, and thus may be evaluated by analytically solving Hamilton's equations.  The outer exponentials may be evaluated by applying the definition of the Lie exponential operator.  Doing so gives

\begin{center}
$\exp(:\pm\displaystyle\int_{0}^{y} a_y dy':)  :
\begin{cases}
p_y \rightarrow p_y \pm a_y       \\
p_x \rightarrow p_x \pm \frac{\partial}{\partial x}\displaystyle\int_{0}^{y} a_y dy
\end{cases}$
\end{center}
\vspace{2mm}
\noindent with all other variables being unchanged.  Furthermore, evaluating $I_{z}$ by applying the definition of the Lie exponential gives 

\vspace{2mm}
\begin{center}
$\exp(:\Delta s a_z:)  :
\begin{cases}
p_y \rightarrow p_y  + \Delta s \frac{\partial a_z}{\partial y}       \\
p_x \rightarrow p_x + \Delta s \frac{\partial a_z}{\partial x}
\end{cases}$
\end{center}
\vspace{2mm}

\noindent with all other variables being unchanged.

For our model, all possible terms are already accounted for in the Bmad symplectic integrator except for the constant term.  Thus, in order to make the appropriate changes to the program, we calculated the normalized vector potential for the constant term in the $\mathbf{B}$-field series and substituted this into our previous evaluations of the Lie exponentials.  Thus for the $b_y$ term we have that
   
\hspace{4.5cm} $\exp(:\pm\displaystyle\int_{0}^{y} a_y dy':)  :
\begin{cases}
p_y \rightarrow p_y      \\
p_x \rightarrow p_x 
\end{cases}$,

\hspace{4.5cm} $\exp(:\Delta s a_z:)  :
\begin{cases}
p_y \rightarrow p_y     \\
p_x \rightarrow p_x - \Delta s\frac{q}{P_0 c}b_y 
\end{cases}$.\\

\noindent We note that in addition to carrying out the aforementioned symplectic integration scheme, the Bmad symplectic integrator also calculates the linear transfer map (that is, the Jacobian) associated with each transformation in equation (7).  From the linear transfer maps we can calculate linear optics parameters, in particular the $\beta$-function, dispersion, and phase advance, at arbitrary locations within a magnet component.  Only the linear transfer maps corresponding to Lie exponentials involving the constant term in the vector potential series, namely, those Lie exponentials we just evaluated, required any changes.  The modifications to the code were straightforward. Taking the Jacobian of the previous Lie exponentials yields the identity matrix in all cases. 

\section{Particle Tracking}
\subsection{Methods}
After making the proper changes to the Bmad symplectic integrator, we were able to begin particle tracking simulations.  We began with a simulation lattice containing a stack representation of the ideal LGB bending magnetic field mentioned in the introduction.  In order to fairly compare the differences between tracking through the map models for each LGB and tracking through a stack, we developed individual simulation lattices containing stack representations of the bending fields for both the C-shaped and canted solenoid LGB prototypes.  For both LGB's, our general method was to track particles through a lattice containing a stack representation, copy the lattice, switch the stack representation for our map model, and repeat the simulations.  Thus, we had four lattices in total, since we had two LGB's, and two models for each LGB.  

\vspace{0.25cm}
In the process of switching out the stack representations for map models, we had to make modifications to the lattices.  In particular, we isolated the LGB's and appended patch elements to either end of the map models.  Patch elements are not physics and exist in simulation lattices only.  Their purpose is to change adjust the reference trajectory and time \cite{bmad} within the lattice.  In our simulations, the patch elements were oriented so that the simulated particle would displace an angle of $2\pi$ upon completing one revolution in the complete lattice.  Because the bending fields of the physical LGB prototypes which we modeled were not designed with the lattice which we began with in mind, we had to scale the fields of the map models.  To do this we simulated initially on-axis particles through the \textit{patch-map model-patch} system until particles which began on-axis, ended on-axis (to 5 digits of precision).  To demonstrate that we scaled the field correctly, we tracked particles with various initial $x,y$ coordinates through both isolated stack representations and map models.  The results are plotted in Figure 6.  As expected, particles with the same initial coordinates have very similar final coordinates, whether they were tracked through a stack or a map model.  After performing this calibration, we then adjusted the supporting quadrupole and bend magnets in the lattices containing map models in order to match the tunes and symmetry of the lattices containing the stacks and to minimize emittance.  

\vspace{0.25cm}
Next we proceeded to track particles through each of our four lattices and produced phase plots of the results.  In tracking simulations for particles with initial $x$ and $y$ values near ($x = y = 50$ $\mu m$) the reference axis, it was expected that the stack would perform in a very similar manner as the map model for both LGB prototypes.  This is because on-axis, the bending field is represented equally well by both the stack and the map model.  However, for simulations of particles with initial $x$ and $y$ values far from the reference axis ($x = y = 1$ $mm$), it was expected that the stack and map model would behave differently.  This is because far off-axis, while they both have the same bending field, the stack and the map model may have very different $B_x$ and $B_z$ components.  To investigate these differences, we carried out such simulations.  For all simulations involving the map model, we determined the number of steps used in our integration scheme in a qualitative manner.  Namely, we increased the number of steps until doing so no longer changed the appearance of the phase plots.  In all cases, this occurred around 2000 steps.  All phase plots depict single particle turn-by-turn tracking, meaning that the simulated particle's data was recorded every turn as it passed a fixed point in the ring.  As many turns were used as were needed to see the qualitative behavior of the phase plots. \\ 

\begin{figure}[H]
\centering
\caption{Comparison of particle tracking through the isolated stack representations and map models for each LGB prototype.}
 \makebox[\textwidth]{\includegraphics[width=7.75in]{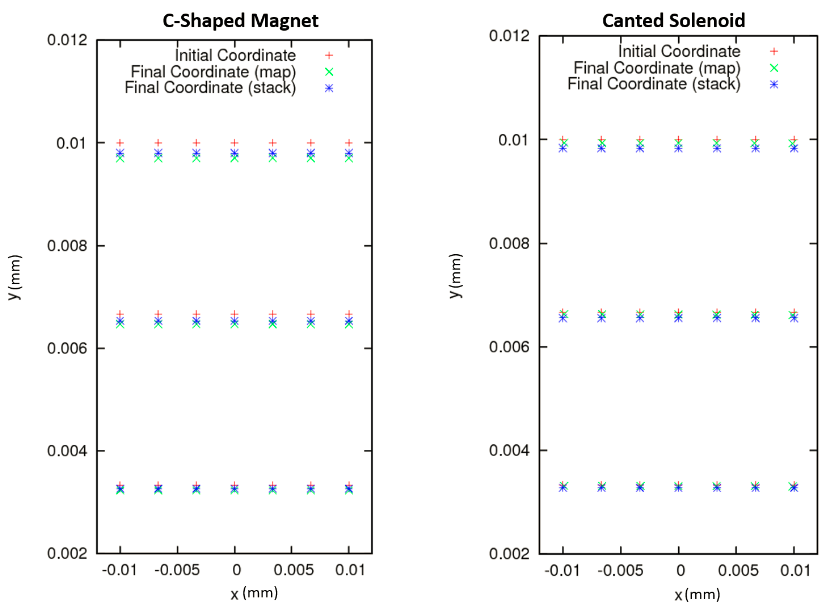}} 
\end{figure}

\subsection{Tracking Results}
Figure 7 depicts phase plots from our particle tracking simulations.  First, it is worth noting that for all instances in which we tracked through the map model, the trajectories in our phase plots do not spiral in or out.  This indicates that we have indeed successfully integrated through the map model symplectically.  

\vspace{0.25cm}
In the case of both LGB prototypes, for particles which began nearly on-axis, there are indeed differences between tracking through the map models and the stack representations.  However, as expected, they are not drastic.  The trajectories through phase space are nearly the same shape and size, and mostly differ in how the data points are clumped together.  On the other hand, for particles which began far off-axis, there is a significant difference between tracking through stack representations as opposed to the map models of each LGB prototype.  The most striking differences are between the $p_y$ vs. $y$ plots.  In particular, for the C-shaped LGB, the map model $p_y$ vs. $y$ plot depicts a ``wavy" ellipse, while the stack representation does not depict this behavior.  Looking to the phase plots for the canted solenoid LGB, in the case of the map model, the $p_y$ vs. $y$ plot depicts a trajectory which changes size and shape.  The stack representation on the other hand, does not demonstrate this.  In the case of both the C-shaped and canted solenoid LGB's, far off-axis it appears that the $p_y$ vs. $y$ plots for the stack representations depict trajectories which are similar to those for the map models, but less sharp.

\setcounter{figure}{6}    
\begin{figure}[H]
\centering
\caption{Phase space plots for particle tracking simulations through the stack representations and map models for both LGB's.}
\subfloat[C-Shaped LGB Tracking Simulations, Particle Initially Nearly on-Axis]{
 \makebox[\textwidth]{\includegraphics[width=7.75in]{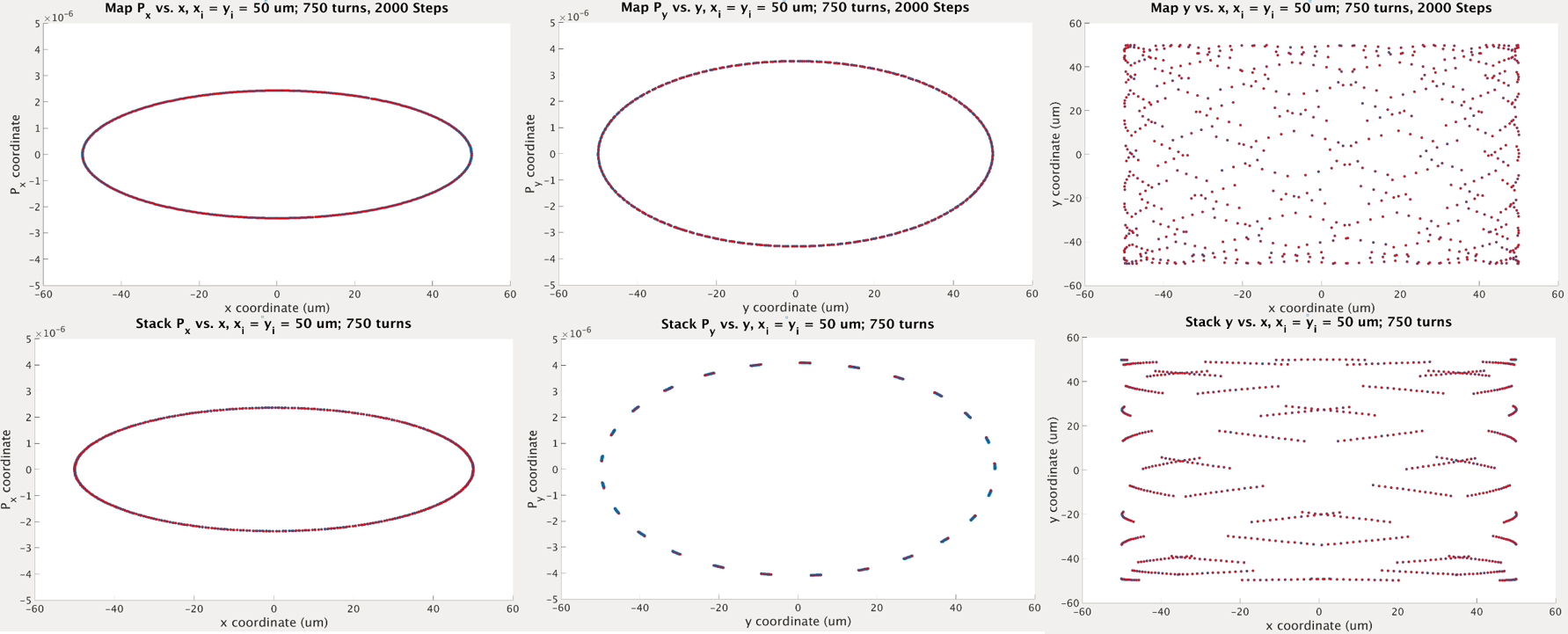}} 
}
\newline
\subfloat[C-Shaped LGB Tracking Simulations, Particle Initially Far off-Axis]{
 \makebox[\textwidth]{\includegraphics[width=7.75in]{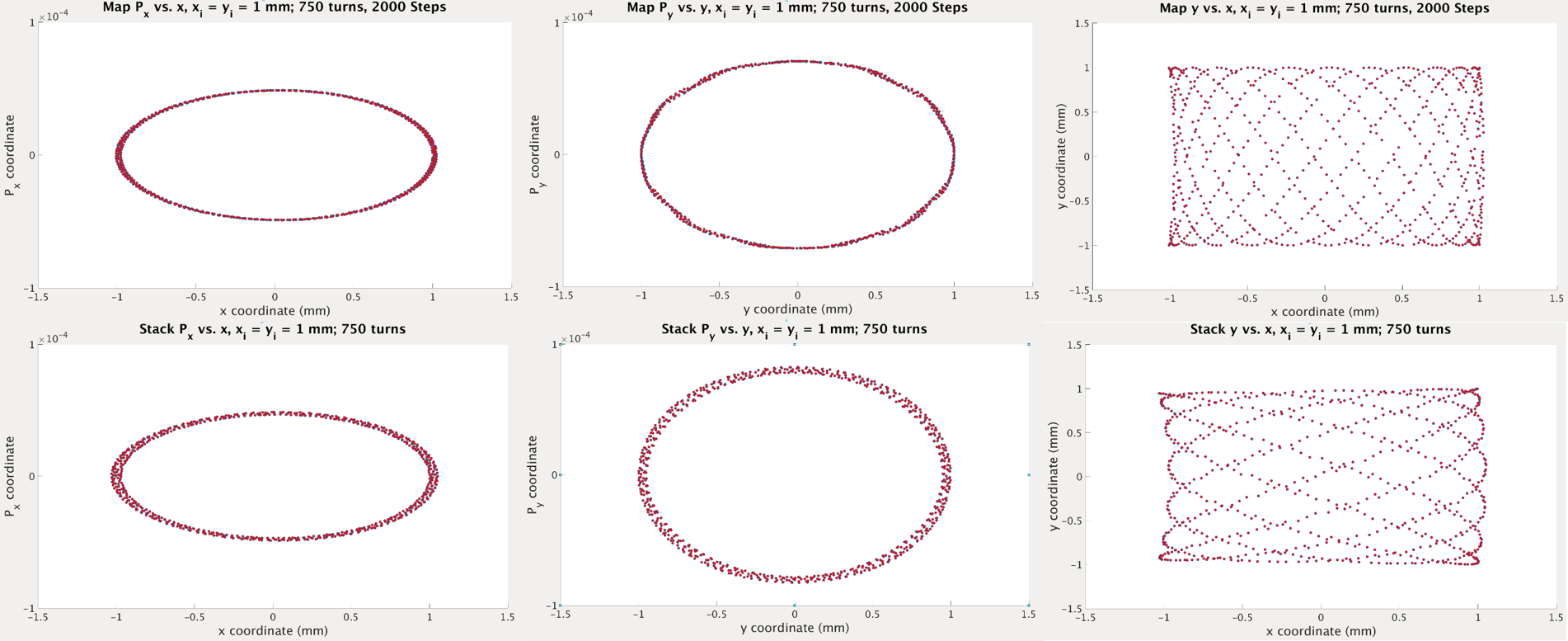}} 
}
\end{figure}
\newpage

\begin{figure}[H]
\ContinuedFloat
\centering
\subfloat[Canted Solenoid LGB Tracking Simulations, Particle Initially Nearly on-Axis]{
 \makebox[\textwidth]{\includegraphics[width=7.75in]{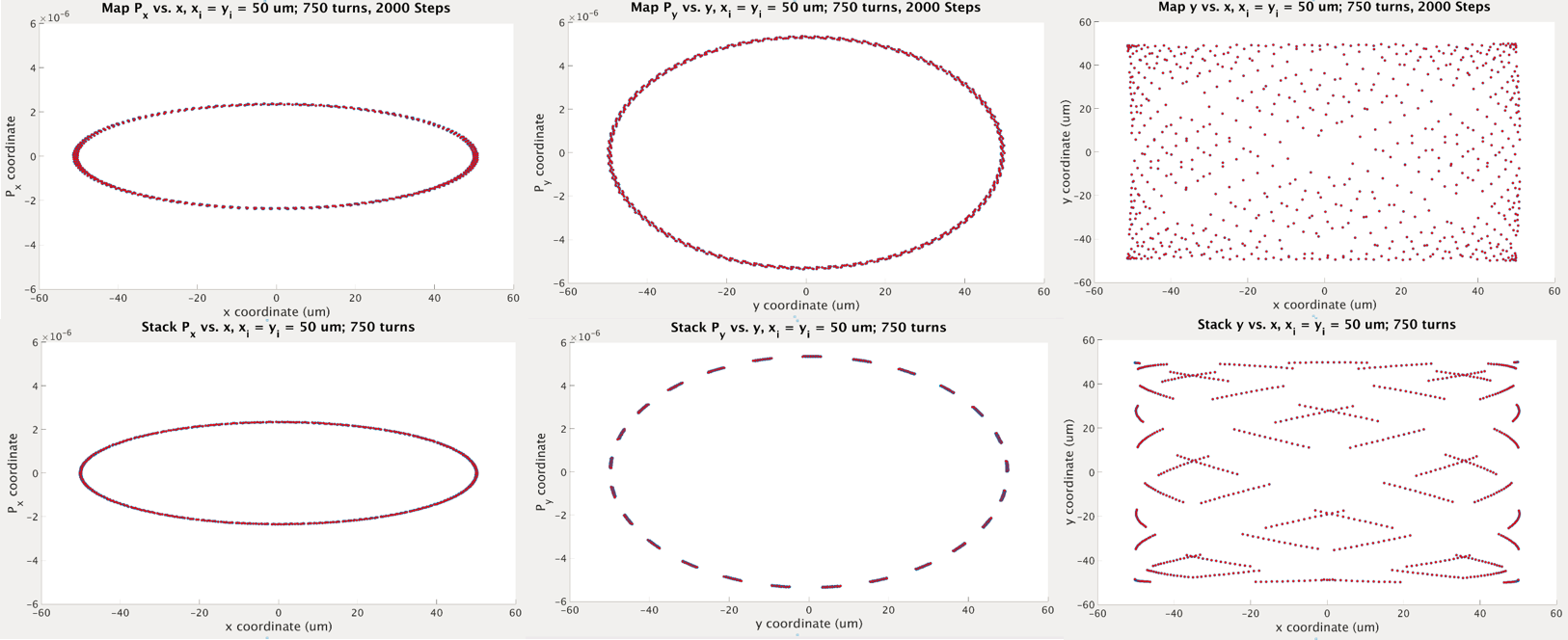}} 
}
\newline
\subfloat[Canted Solenoid LGB Tracking Simulations, Particle Initially Far off-Axis]{
 \makebox[\textwidth]{\includegraphics[width=7.75in]{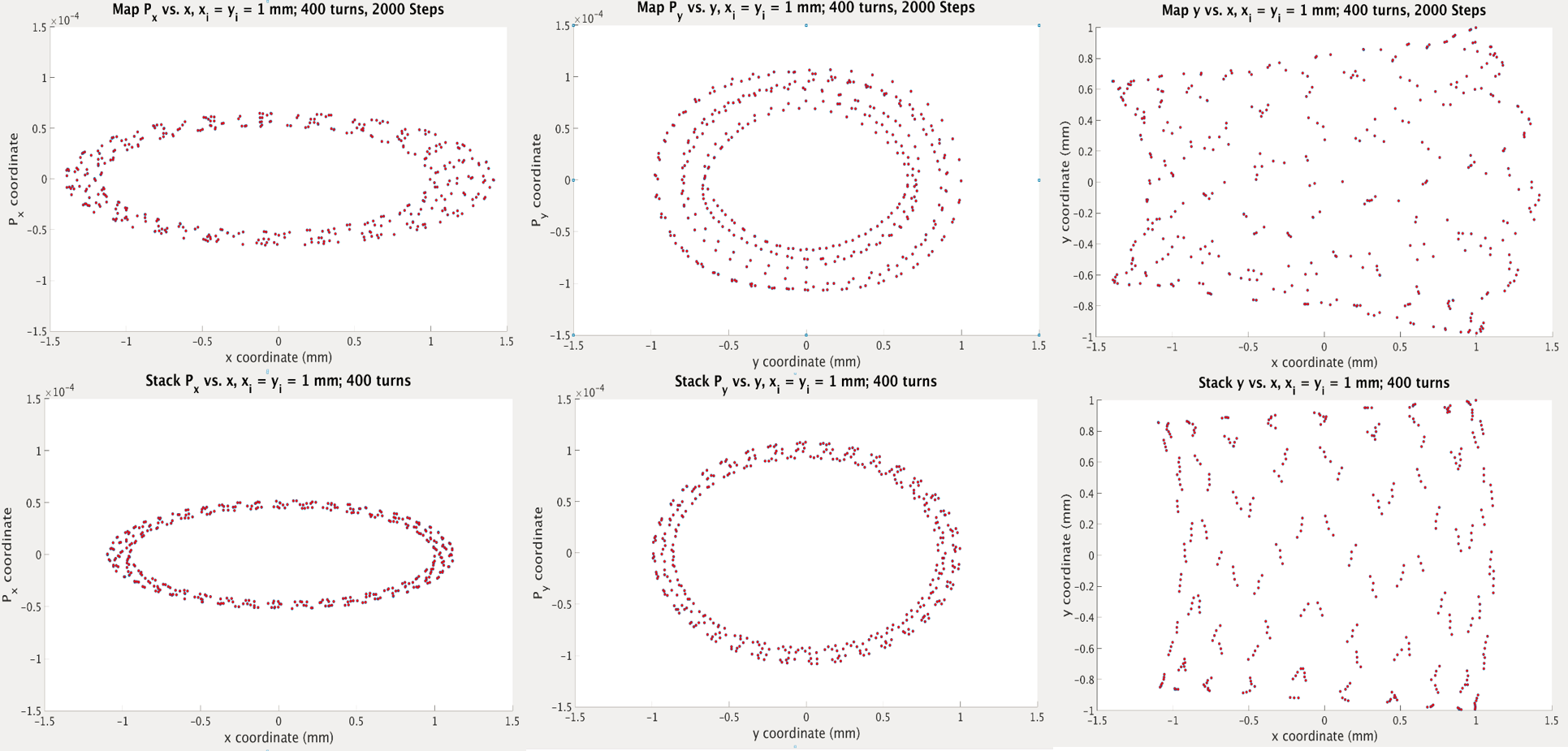}} 
}

\end{figure}

\subsection{Speed Comparisons}
We end our results section with a few words on the relative speeds of tracking through lattices containing the map models as opposed to tracking through the stack representations.  Table 4 provides a comparison between the two, giving the time to run a simulation in each case.  We recorded the time for $2000$ steps because this was the minimum number for which adding any more steps did not increase the accuracy of our simulations, that is, did not yield a significant difference in the phase space plots.  Table 4 also provides the number of steps needed so that the map model simulations could be carried out in the same time as the stack representation simulations.  \\


\begin{table}[H]    
\caption{Comparison of speeds for tracking through lattices containing map models and stack representations} 
\centering
\begin{tabular}{|c|c|}

  \hline
  Time for 2,000 Steps, C-Shaped LGB & 3 s   \\ 
  \hline
   Time for C-Shaped LGB Stack & 0.5 s \\
  \hline
  Steps for Equal Time, C-Shaped LGB & 100 steps  \\
  \hline
  Time for 2,000 Steps, Canted Solenoid LGB & 3 s   \\ 
  \hline
   Time for Canted Solenoid LGB Stack & 0.5 s \\
  \hline
  Steps for Equal Time, Canted Solenoid LGB & 100 steps  \\
\hline

\end{tabular}
\end{table}

\section{Conclusions}
By fitting to data with a functional series, we were able to develop a facility for the SLS2 project which allows for linear optics calculations and tracking studies through storage rings containing components with arbitrary magnetic field profiles.  With this model, we were successful in representing both the C-shaped and canted solenoid LGB's.  After developing lattices containing stack representations for the LGB prototypes,  we were able to replace the stacks with the map models.  Then, after we adjusted the quadrupoles and bends of the lattices, we were able to successfully track through them. In addition, the stability of our phase plot trajectories demonstrates that we able to track through the map model symplectically, validating our changes to the existing Bmad symplectic integrator.  Furthermore, we found that tracking through lattices containing a map model is reasonably fast, requiring $~3$ seconds for 2000 steps, whereas it takes $~0.5$ seconds to track through a stack representation.  Moving forward, the procedure outlined in this report will be repeated with higher accuracy map models produced by using more terms in our series in equation (4).

\newpage
\section{Appendix}
Table 1 lists the fit parameters for the C--shaped and canted solenoid LGB's.  Note that in each case, $b_y$ is not listed explicitly.  Rather, it corresponds to $C$ in the case that all other parameters in the row are zero.  Any fit parameter from the model derived in Section 2 which is not stated is zero. 

\setcounter{table}{0}
\begin{table}[H]
\caption{Fit Parameters for the C--shaped and canted solenoid LGB's}

\vspace{5mm}
\begin{tabular}{|l|l|l|l|}
\hline
\multicolumn{4}{|c|}{C-Shaped LGB}  \\ \hline

Term number    &$C_n$                                    &  $k_{xn}$                                  &      $ k_{zn}$                 \\ \hline
1                            &32.21300273                          & 0                                              & 0                        \\ \hline
2                            &-0.5738766143 &0 &-32.20025548   \\ \hline
3                            &-13.3176724   &0 &-11.92544178     \\ \hline
4                            &-16.49834507  &0 &-11.94797965 \\ \hline
5                            &-64.89878143  &0 &-4.996082992    \\ \hline
6                            &28.45760474   &0 &-9.480756703        \\  \hline
7                            &29.13820758   &0 &-9.473793469     \\ \hline

\multicolumn{4}{|c|}{Canted Solenoid LGB}  \\ \hline

Term number    &$C_n$                                    &  $k_{xn}$                                  &      $ k_{zn}$                 \\ \hline
1                            &-2.4111959789306           & 0                                              & 0                                               \\ \hline
2                            &-0.0623891894555       &-102.0440741217437              &104.1185795259480  \\ \hline
3                            &-0.0230512693537       &-140.0630090173596              &122.0433760269421   \\ \hline
4                            &-0.0061954195378       &-202.8245989448726              &139.3502247105487   \\ \hline
5                            &-2.4710095044701       &-11.3595060233678              &14.1759090008411   \\ \hline
6                            &-0.9909810125361       &-25.2608193368136              &32.1060892779629       \\  \hline
7                            &-0.5249565453637       &-40.1332078703115              &50.1715421129154    \\ \hline
8                            &-0.1411708088290       &-76.2423493478569              &86.1044550658385     \\ \hline
9                            &-0.2843127125471       &-56.3705209191432              &68.1229211975367    \\ \hline

\end{tabular}
\end{table}

\end{document}